\documentclass[aps,pra,amsmath,amssymb,amsfonts,superscriptaddress,floatfix]{revtex4} 
\pdfoutput=1
\usepackage{amsmath,amsthm,amsfonts,amssymb}
\usepackage{graphicx}

\usepackage{color}
\usepackage{hyperref}

\usepackage{algorithm}
\usepackage[capitalize,nameinlink]{cleveref}

\newcommand{\be}{\begin{equation}}
\newcommand{\ee}{\end{equation}}

\begin{document}

\title{A Personal History of the Hastings-Michalakis Proof of Hall Conductance Quantization}

\begin{abstract}
This is
a personal history of the Hastings-Michalakis proof of quantum Hall conductance quantization.
\end{abstract}

\author{Matthew B.~Hastings}
\maketitle

The Hall conductance quantization was an open problem in mathematical physics for a long time.  A fundamental physical argument for the quantization was given by Laughlin\cite{laughlin1981quantized}, but finding a mathematical proof for an interacting system of electrons remained open until \cite{hastings2015quantization}.  
In this note, I give the history of this proof in brief.  Recently, a pair of opinion pieces appeared in Nature and Nature Reviews Physics discussing this specific paper.   Since those pieces do not accurately convey the history, it may be of some interest to give the broader history, emphasizing what new mathematical tools and ideas were needed and why those tools and ideas were developed.  The focus will be on the proof of \cite{hastings2015quantization} itself, so references will unfortunately be incomplete.

Of course, Hall conductance quantization had been proven in various forms before.
The result was proven for free fermions\cite{bellissard1994noncommutative} using noncommutative geometry techniques and had been proven for interacting systems by Avron and Seiler under an additional {\it averaging} assumption\cite{avron1985quantization}, but 
there was no proof for an interacting system without this averaging assumption.  This elegant averaging proof was extremely important for the Hall effect proof of \cite{hastings2015quantization}; roughly, the Hall effect proof of \cite{hastings2015quantization} involved replacing the average curvature of a connection due to adiabatic evolution as considered in \cite{avron1985quantization} with the curvature of a quasi-adiabatic evolution operator, as I'll explain more below.

The story of the proof of Hall conductance quantization proof of \cite{hastings2015quantization} starts, for me, years earlier around 2002-2003.  
It starts with the proof of the higher dimensional Lieb-Schultz-Mattis theorem (LSM)\cite{hastings2004lieb}, which is where the tools needed to prove Hall conductance quantization were developed.  I had read a very insightful article by Misguich and Lhuiller\cite{misguich2000some} on why they believed such an LSM theorem should exist; up to that point it had only been proven in one-dimension.  
This article was an inspiration to find a proof of such an LSM theorem, and I developed the tool of quasi-adiabatic continuation to do this. 

In brief, the Lieb-Schultz-Mattis theorem in one dimension proves that a system of spin-$1/2$ particles with an $SU(2)$ symmetric Hamiltonian and translation invariance cannot have a unique ground state with a spectral gap.  Further extensions with Affleck\cite{affleck1986proof} generalized the theorem and clarified the implications of the theorem.  Roughly, these one-dimensional spin systems must either break some discrete symmetry (in this case, translation symmetry), in which case the ground state becomes degenerate or close to degenerate; or the system may break a continuous symmetry (such as ferromagnetically ordering), in which case there is a continuous excitation spectrum; or the system may form a one-dimensional spin liquid with spinon excitations, again having a continuous excitation spectrum.
Attempts to extend the theorem beyond one dimension\cite{affleck1988spin} ran into trouble.  In a sense, the variational argument of Lieb, Schultz, and Mattis was too ``violent" in its effects on the ground state.

Misguisch and Lhuillier however argued that such a theorem should hold in higher dimensions, noting that in addition to the above possibilities, the system could 
have some kind of topological order, bringing a beautiful connection to topology.   Indeed, in that case the system would still be unable to have a unique gapped ground state but for a different reason: there would be a topological degeneracy.

The tool of quasi-adiabatic continuation introduced in \cite{hastings2004lieb} to prove the LSM theorem was more ``gentle" than previous variational attempts, and was able to diagnose the presence of topological order.
This tool served as a general way to get precise theorems out of physical arguments involving slowly inserting gauge flux into some physical system; it is worth mentioning the very important work of Oshikawa here too\cite{oshikawa2000commensurability} who had the idea that the Lieb-Schultz-Mattis theorem might be related to  flux insertion.                                                                                               
Misguich and Lhuillier realized that on physical grounds one might expect the gap to not depend much on 
flux
for a liquid-like state (though counter-examples involving first order phase transitions at particular values of the 
flux
can be constructed) making flux insertion seem indeed like a good way to generate variational states; quasi-adiabatic continuation allowed one to make this idea precise so that even if the gap did
close one could ``pretend" it did not for some purposes.

Quasi-adiabatic continuation allowed one to show that, given a gapped local Hamiltonian, the change in the ground state under a small change in the Hamiltonian could be given by a {\it local} and Hermitian operator acting on the ground state but much of the use of this technique is to apply this operator even in a case where the gap might close.
This operator played a key role in the Hall proof.

The proof of the LSM theorem in higher dimensions\cite{hastings2004lieb} introduced other tools in addition to quasi-adiabatic continuation, tools that would be needed for Hall conductance quantization.  This paper re-discovered Lieb-Robinson bounds, and gave the first proof of these bounds in a way that was independent of local Hilbert space dimension.  This alternative proof, which at the time might have seemed merely to improve some constants since the local Hilbert space dimension in those applications typically was small ($2$ or $4$), in fact turned out to be an essential ingredient in the Hall conductance proof.  This alternative proof method has also become a general template for proving many variations of Lieb-Robinson bounds such as in systems with both short and long range interactions or systems with approximately commuting terms in the Hamiltonian.

The Lieb-Robinson bounds generally show how excitations can spread in a lattice system, showing that there is an {\it approximate} light-cone in these systems limiting how fast the system can respond to any perturbation.  Also, this paper \cite{hastings2004lieb} proved the exponential decay of correlations in gapped lattice systems with a local Hamiltonian; this proof is an example of how many of the proofs involving Lieb-Robinson bounds in gapped systems go.  First, one uses some analytic technique relying on the spectral gap to relate some quantity (in this case, a correlation function) to a commutator (in this case, a Green's function given by an expectation value of a commutator) and then one uses Lieb-Robinson bounds to control that commutator.

 It soon became apparent to me that this general tool of quasi-adiabatic continuation could be used to make Hall conductance quantization arguments precise and to remove the averaging assumption.  Rather than needing a single flux insertion as in the LSM theorem which serves to detect a topologically ordered ground state, one would need two flux insertions to detect a curvature in the response of the ground state.
  It was clear to me that the topological arguments of Thouless and collaborators\cite{thouless1982quantized} could be made  general using this tool.  These arguments of Thouless et. al. served as a precursor to the averaging argument of Avron and Seiler\cite{avron1985quantization}.  To handle a many-body system without averaging assumptions,
 one would need to compute the curvature of evolution under this quasi-adiabatic continuation operator.

At a technical level, the LSM theorem relied on a ``virtual flux", and it was also clear that the Hall conductance proof would then need two virtual fluxes.  The virtual flux plays the following role.  One considers a system on a torus, and one wishes to understand the response of the system to some flux.  Here we
imagine Aharonov-Bohm fluxes threading
one or two directions of the torus.  To give a picture, let us regard the torus as a square with opposite sides identified; introduce $x,y$ coordinates, periodic with periods $L_x,L_y$.  By making a gauge choice, one can consider this flux as being implemented by a gauge field which is nonvanishing on some vertical line such as $x=0$ (or similarly, a flux may be inserted on some horizontal line).  In general, the ground state $\Psi_0$ could have some very complicated response to the flux, but one wishes to show that there is {\it some} state $\Psi_{\theta}$ which is an approximate eigenstate of the Hamiltonian with a flux $\theta$ inserted that has the following two properties: far from the line $x=0$, the reduced density matrix of $\Psi_\theta$ is (almost) the same as that of $\Psi_0$, while near the line $x=0$, the reduced density matrix of $\Psi_\theta$ is (almost) related to that of $\Psi_0$ by some gauge transformation by angle $\theta$.

To construct $\Psi_{\theta}$, we use quasi-adiabatic continuation; while quasi-adiabatic continuation is defined to match adiabatic evolution in the case that the ground state has a spectral gap, we apply quasi-adiabatic continuation even if the gap closes!  (If the gap closes, the state produced by quasi-adiabatic continuation might not be the ground state but it still may be useful as a variational state or for other purposes.)  Quasi-adiabatic continuation changes the state only near the line, giving the first property immediately.  To get the second property, imagine also inserting an opposite flux along the line $x=L_x/2$; then, by locality of quasi-adiabatic continuation, this second insertion has no effect near $x=0$ (so, even if the insertion near $x=L_x/2$ is not done, one may {\it pretend} it is done for the purposes of computing properties near $x=0$), but the combined effect of the two insertions is a gauge transformation since the net flux is zero.

This property of the state produced by quasi-adiabatic continuation, that locally it is related to the ground state by a gauge transformation, is related to the ``gentleness" of this method and it is why the curvature of the quasi-adiabatic continuation operator could be used to prove the Hall conductance quantization.  The averaging method of Avron and Seiler for Hall conductance considered the curvature of {\it adiabatic} evolution as a function of two fluxes, $\theta,\phi$, defining a so-called ``flux torus".  The curvature at $\theta=\phi=0$ could be related to the Hall conductance and the {\it average} of the curvature could be proven to be quantized by topological methods.  Using the quasi-adiabatic continuation operator instead, this ``gentleness" meant that the average of the curvature of the quasi-adiabatic evolution operator indeed was (almost) equal to the curvature at $\theta=\phi=0$; indeed, the curvature depended only weakly on flux angle in that case.

The conceptual ingredients were then all in place to prove the Hall conductance quantization.  However, these kinds of proofs tended to be rather lengthy, involving an enormous number of triangle inequalities and optimization over many parameters.  The Hall conductance proof promised to be even more lengthy.  One problem is that one needed to show that evolution under this quasi-adiabatic evolution operator also had locality properties, i.e., that this operator also had a Lieb-Robinson bound (here is where the dimension independent Lieb-Robinson bounds were necessary).
  I was at Los Alamos National Laboratory at the time and I chanced after that to receive a grant to hire a postdoc and I had an applicant, Spiros Michalakis, coming from a mathematical physics group.  So, it seemed like he would be good to work with on finishing the detailed estimates.  We succeeded, and we found ultimately a rather clean and simple proof.  
  The key ingredient in the proof indeed was quasi-adiabatic continuation.
  
  One key to making a clean proof was to use a modified ``exact" form of quasi-adiabatic continuation, 
originally introduced by Tobias Osborne\cite{osborne2007simulating} in 2007, a few years after \cite{hastings2004lieb}; Osborne realized that some of the approximations made in the quasi-adiabatic continuation operator in the LSM theorem could be made exact at the cost of turning some other exponentially small errors into merely super-polynomially small errors, and that the needed Lieb-Robinson bound for this operator held.  I realized that an old result in analysis\cite{ingham1934note} showed that these errors could be made ``almost exponentially small" so that one could still find fairly tight bounds in this way.

\bibliographystyle{unsrturl}                                                                                            \bibliography{qh-ref}   
\end{document}